\documentclass[%
 reprint,
 amsmath,amssymb,
 aps,
 prl,
]{revtex4-1}

\usepackage{graphicx}
\usepackage{dcolumn}
\usepackage{bm}
\usepackage{ulem}
\usepackage{color}

\newcommand{\nc}{\newcommand}
\nc{\bra}{\langle}
\nc{\ket}{\rangle}
\nc{\vac}{|0\ket}
\nc{\da}{^{\dagger}}
\nc{\ps}{\hat{\psi}}
\nc{\pd}{\hat{\psi\da}}
\nc{\HASEP}{\hat{\mathcal{H}}_{\text{ASEP}}}
\nc{\Hboson}{\hat{\mathcal{H}}_{\text{boson}}}
\nc{\HDNLS}{\hat{\mathcal{H}}_{\text{DNLS}}}
\nc{\red}{\textcolor{red}}
\nc{\sred}[1]{\textcolor{red}{\sout{#1}}}

\begin{document}


\title{Burgers equation with finite particle correction of the asymmetric simple exclusion process derived from the derivative nonlinear Schr\"{o}dinger equation}

\author{Yuki Ishiguro$^1$, Jun Sato$^2$ and Katsuhiro Nishinari$^3$}
 \affiliation{$^1$Department of Aeronautics and Astronautics,Faculty of Engineering,University of Tokyo,Hongo,Bunkyo-ku,Tokyo 113-8656,Japan\\
$^2$Physics Department and Soft Matter Center, Ochanomizu University, Japan\\
$^3$Research Center for Advanced Science and Technology, University of Tokyo,4-6-1 Komaba,Meguro-ku,Tokyo 153-8904,Japan}

\begin{abstract}
We investigate the dynamics of the asymmetric simple exclusion process (ASEP) on a ring. 
The ASEP is equivalent to the derivative nonlinear Schr\"{o}dinger equation (DNLS), which is integrable quantum field theory, in the continuous limit. 
We derive the Burgers equation with finite particle correction from the DNLS and 
numerically confirm that the obtained Burgers equation describes the dynamics of the ASEP at small numbers of particles better than the conventional Burgers equation.
\end{abstract}

\maketitle

\begin{figure}[h]
    \centering
    \includegraphics[height=2.5cm]{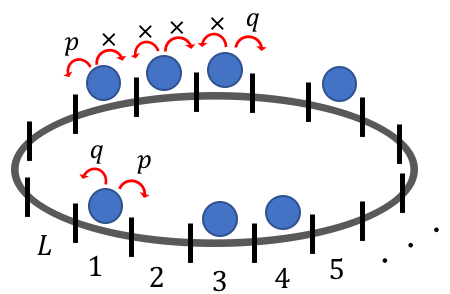}
    \caption{ASEP with a periodic boundary}
    \label{fig:ASEP}
\end{figure}

\textit{Introduction.}---
Although almost all phenomena in the world are out of equilibrium, the theory of nonequilibrium physics remains incomplete.
Therefore, the investigation of universal laws among nonequilibrium phenomena is important work in modern theoretical physics.
The asymmetric simple exclusion process (ASEP), which is a continuous time Markov model describing asymmetric diffusion of hard-core particles in one dimension, is among the models holding most hope for studying nonequilibrium transport phenomena \cite{derrida,phase1,phase2,KPZ,KPZ2}.
A schematic drawing of the ASEP is shown in Fig.\ref{fig:ASEP}.
The update rule of this model is simple. Each particle hops to the nearest right (left) site with probability $pdt$ ($qdt$) in the time interval $dt$ if that site is vacant. If another particle already exists in that site, hoppings do not occur. 
Despite its simplicity, the ASEP describes various interesting nonequilibrium phenomena, such as the boundary-induced phase transition \cite{phase1,phase2}.
Moreover, current fluctuation is known to belong to Kardar-Parisi-Zhang universality class to which surface growth phenomena belong \cite{KPZ,KPZ2}.
The ASEP was originally introduced as a model of biopolymerization \cite{bio}. Then it was applied to a wide range of nonequilibrium phenomena, such 
as traffic flow, pedestrian dynamics \cite{traffic,traffic2}, and biophysical transport \cite{biotransport,biotransport2}.
In addition to these wide applications in nonequilibrium physics, the mathematical aspect of the ASEP also fascinates many researchers.
The ASEP is a quantum integrable system, and its Hamiltonian is exactly diagonalized by the Bethe ansatz \cite{bethe,bethe2}. 
If an initial state can be expressed by the superposition of Bethe vectors, then its dynamics can be calculated exactly.
Moreover, the exact stationary state is constructed from the matrix product ansatz \cite{solvers}.

In this letter, we investigate the mathematical structure of the ASEP. 
It is known that the ASEP is related to the two types of integrable systems: the Burgers equation, which is a classical integrable system \cite{burgers1,burgers2}, and the derivative nonlinear Schr\"{o}dinger equation (DNLS), which is an integrable quantum field theory \cite{DNLS}.
It has been shown that the Burgers equation is derived from the ASEP by taking the mean field approximation and the continuous limit.
On the other hand, the DNLS is obtained via the bosonization of the ASEP and the continuous limit \cite{BosonModel}.
However, the relation between the Burgers equation and the DNLS has not been uncovered.
Therefore, we investigated this relation in detail and found the derivation of the Burgers equation from the DNLS.
Moreover, the obtained Burgers equation contains finite particle correction of the ASEP and describes the dynamics of the ASEP at low particles better than the conventional Burgers equation.

\textit{Formulation of the ASEP.}---
Let us consider the ASEP on a ring. The number of lattices is $L$.
We introduce the variable $n_j$ as the number of particles at site $j$. 
In the case of the ASEP, $n_j$ is $0$ or $1$ due to the exclusive volume effect.
A local state at site $j$ is represented by the local basis $(|1\ket,|0\ket)$ which are identified with the two-dimensional vectors $((0,1)^{\text{T}},(1,0)^{\text{T}})$.
If a particle exists at site $j$, $|n_j\ket =|1\ket$. Otherwise, $|n_j\ket =|0\ket$.
We introduce the orthogonal basis $|n\ket =|n_1,n_2,\cdots,n_L \ket$ which is a tensor product of the local basis. A state of the system at time $t$ is written as
\begin{equation}
    |\psi(t)\ket = \sum_n \psi(n,t)|n\ket,
    \label{eq:state}
\end{equation}
where $\psi(n,t)$ means the probability that the system is in a state $(n)$ at time $t$.
The ladder operators $\hat{s}_{j}^{\pm}$ and the particle number operator $\hat{n}_{j}$  are defined by the Pauli matrices $\hat{\sigma}$ as $\hat{s}_{j}^{\pm}=\frac{1}{2}(\hat{\sigma}_{j}^{x}\pm\hat{\sigma}_{j}^{y})$, $\hat{n}_{j}=\frac{1}{2}( 1-\hat{\sigma}_{j}^{z})$.
The time evolution of this state obeys the master equation
\begin{equation}
    \frac{d}{dt}|\psi(t)\ket = -\HASEP|\psi(t)\ket
    \label{eq:MasterASEP}
\end{equation}
where the Markov matrix $\HASEP$ is given by
\begin{align}
\begin{split}
    \HASEP 
    = \sum_{j=1}^{L} &\left[ - p\hat{s}_{j}^{+}\hat{s}_{j+1}^{-}-q\hat{s}_{j}^{-}\hat{s}_{j+1}^{-} \right. \\
    & \left. + p\hat{n}_{j}(1-\hat{n}_{j+1}) + q(1-\hat{n}_{j})\hat{n}_{j+1} \right].
    \label{eq:HASEP}
\end{split}
\end{align}
This equation has the same form as the imaginary time Schr\"{o}dinger equation, so we call $\HASEP$ as Hamiltonian hereafter. 
We introduce the projection state $\bra s|$ defined as
\begin{equation}
    \bra s|=\bra0|\text{exp}\left( \sum_{j=1}^L \hat{s}_{j}^{+} \right) = \sum_n \bra n|
\end{equation}
where $|0\ket$ denotes the vacuum state $|0\ket=|0,0,\cdots,0\ket$.
Then, the normalization condition of the state vector is given by
\begin{equation}
    \bra s|\psi(t)\ket = 1
\end{equation}
and the expectation value of the physical quantity $\hat{A}$ in a state $|\psi(t)\ket$ is written as
\begin{equation}
    \bra \hat{A} \ket = \bra s|\hat{A}|\psi(t)\ket.
    \label{eq:ExpASEP}
\end{equation}
The time evolution of particle density is obtained from eq. (\ref{eq:MasterASEP}), (\ref{eq:HASEP}), and (\ref{eq:ExpASEP}), and is given by 
\begin{align}
\begin{split}
    \frac{d}{dt}\bra\hat{n}_i\ket 
    &= p\bra\hat{n}_{i-1}\ket+q\bra \hat{n}_{i+1}\ket \\
    &-(p+q)\bra \hat{n}_i\ket+(p-q)\bra \hat{n}_i(\hat{n}_{i+1}-\hat{n}_{i-1})\ket.
    \label{eq:ParticleDenASEP}
\end{split}
\end{align}

\textit{The Burgers equation derived from the ASEP in the continuous limit}---
Before explaining our results, we review the conventional derivation of the Burgers equation from the ASEP.
First, we adopt the mean field approximation, i.e. we ignore the correlations among particles. We replace the two-point correlation function $\bra \hat{n}_i(\hat{n}_{i+1}-\hat{n}_{i-1})\ket$ with $\bra \hat{n}_i \ket \bra \hat{n}_{i+1}-\hat{n}_{i-1}\ket$. Then the time evolution equation of particle density (\ref{eq:ParticleDenASEP}) is written as 
\begin{equation}
\begin{split}
    \frac{d}{dt}\bra\hat{n}_i\ket 
    &= p\bra\hat{n}_{i-1}\ket+q\bra \hat{n}_{i+1}\ket \\
    &-(p+q)\bra \hat{n}_i\ket+(p-q)\bra \hat{n}_i\ket \bra \hat{n}_{i+1}-\hat{n}_{i-1} \ket.
    \label{eq:MeanParticleDenASEP}
\end{split}
\end{equation}
Second, we take the continuous limit. 
We denote the total length of the system as $l$ and the lattice distance as $a=\frac{l}{L}$. 
Then we take the infinite limit of the lattice number $L$. 
We write $\bra \hat{n}_j \ket=\rho(x_j,t)$ and rescale time $at \rightarrow t$, then we obtain the Burgers equation
\begin{equation}
    \partial_t\rho = aD\partial_x^2\rho -2\alpha\rho+4\alpha\rho\partial_x\rho.
    \label{eq:burgers}
\end{equation}
Therefore, the Burgers equation describes the time evolution of particle density of the ASEP in the continuous limit.

\begin{figure}[h]
  \begin{center}
    \begin{tabular}{c}
      \begin{minipage}{0.5\hsize}
        \begin{center}
          \includegraphics[height=2.5cm]{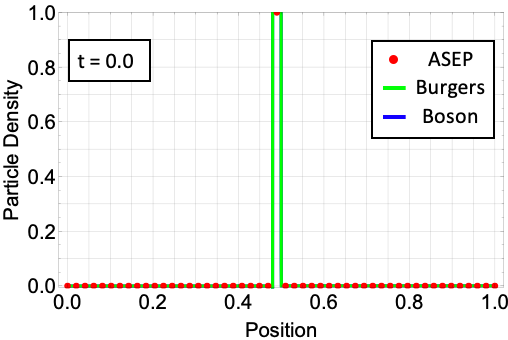}
        \end{center}
      \end{minipage}
      \begin{minipage}{0.5\hsize}
        \begin{center}
          \includegraphics[height=2.5cm]{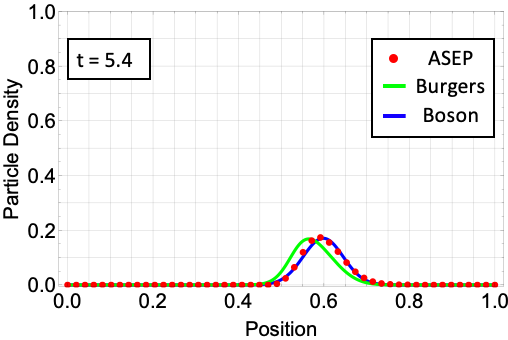}
        \end{center}
      \end{minipage} \\
      
      \begin{minipage}{0.5\hsize}
        \begin{center}
          \includegraphics[height=2.5cm]{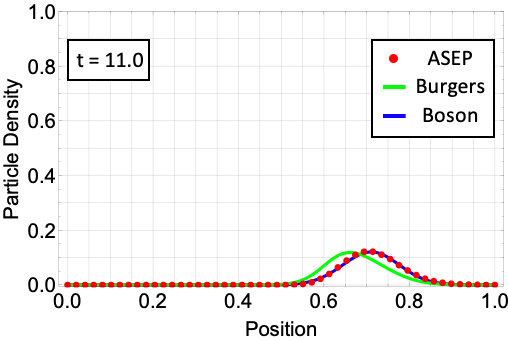}
        \end{center}
      \end{minipage}            
      \begin{minipage}{0.5\hsize}
        \begin{center}
          \includegraphics[height=2.5cm]{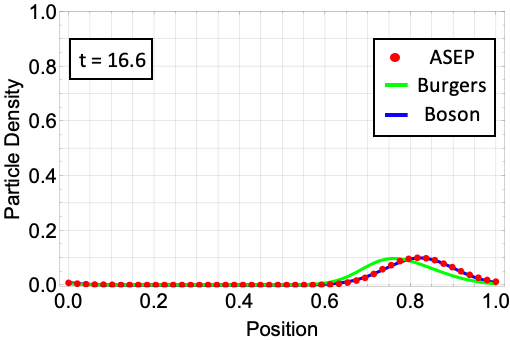}
        \end{center}
      \end{minipage}  \\
      
    \begin{minipage}{0.5\hsize}
        \begin{center}
          \includegraphics[height=2.5cm]{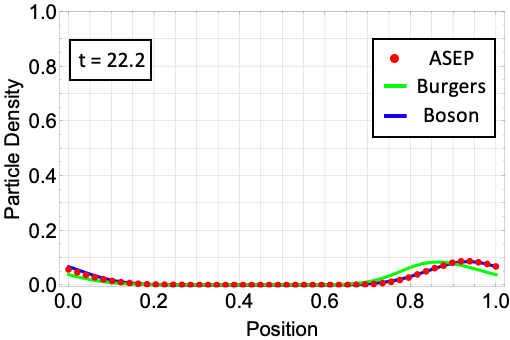}
        \end{center}
      \end{minipage}            
      \begin{minipage}{0.5\hsize}
        \begin{center}
          \includegraphics[height=2.5cm]{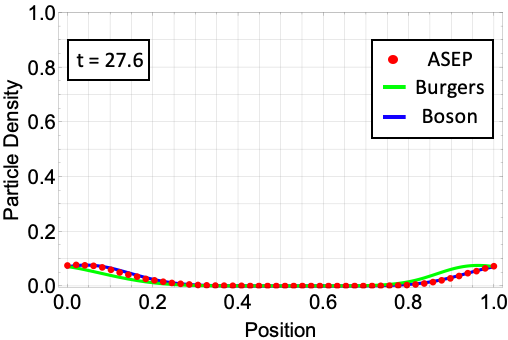}
        \end{center}
      \end{minipage}  \\
      
    \end{tabular}
   \caption{Particle density of the ASEP (red dots), and numerical solutions of the Burgers equation (green lines) and of the Boson-Burgers equation (blue lines) for particle number $N=1$. The localized wave represented by green lines is receding against that of the ASEP represented by red dots. Particle density values along blue lines almost match those of the red dots.}
    \label{fig:CompBurgersASEP}
  \end{center}
\end{figure}

\begin{figure}[h]
  \begin{center}
    \begin{tabular}{c}
      \begin{minipage}{0.5\hsize}
        \begin{center}
          \includegraphics[height=2.5cm]{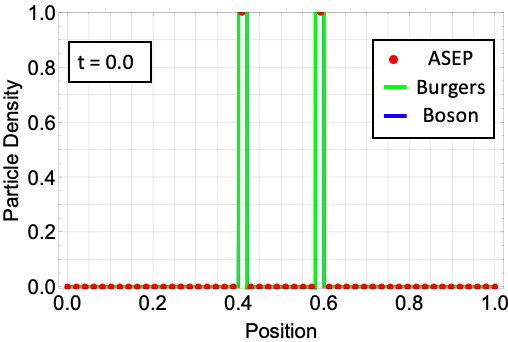}
        \end{center}
      \end{minipage}
      \begin{minipage}{0.5\hsize}
        \begin{center}
          \includegraphics[height=2.5cm]{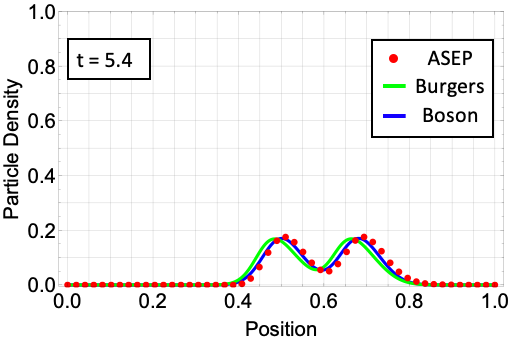}
        \end{center}
      \end{minipage} \\
      
      \begin{minipage}{0.5\hsize}
        \begin{center}
          \includegraphics[height=2.5cm]{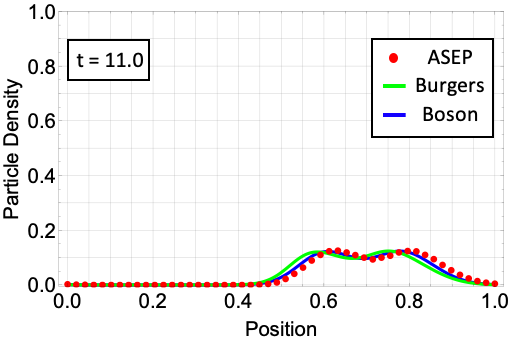}
        \end{center}
      \end{minipage}            
      \begin{minipage}{0.5\hsize}
        \begin{center}
          \includegraphics[height=2.5cm]{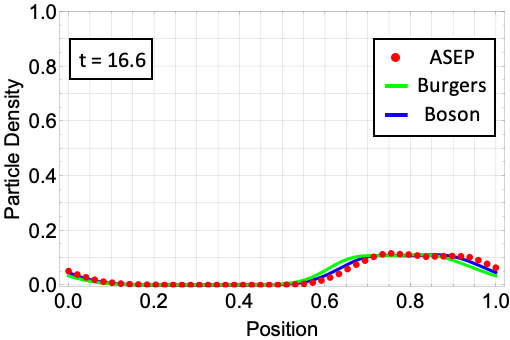}
        \end{center}
      \end{minipage}  \\
      
    \begin{minipage}{0.5\hsize}
        \begin{center}
          \includegraphics[height=2.5cm]{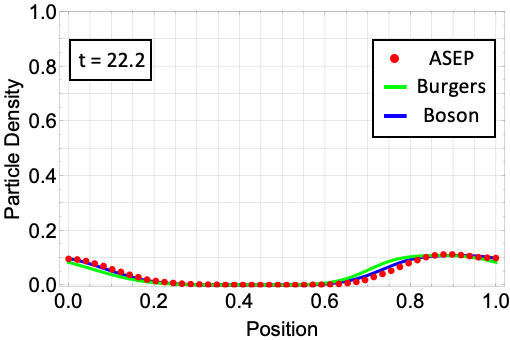}
        \end{center}
      \end{minipage}            
      \begin{minipage}{0.5\hsize}
        \begin{center}
          \includegraphics[height=2.5cm]{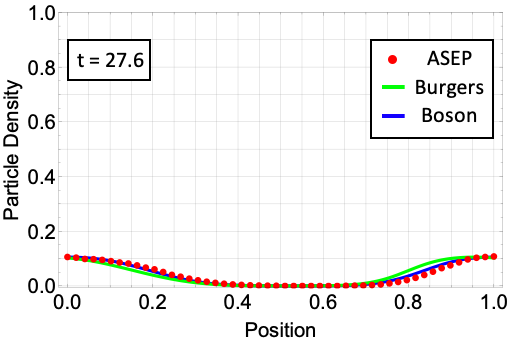}
        \end{center}
      \end{minipage}  \\
      
    \end{tabular}
   \caption{Particle density of the ASEP (red dots), and numerical solutions of the Burgers equation (green lines) and of the Boson-Burgers equation (blue lines) for particle number $N=2$. As in the case of $N=1$, the Boson-Burgers describe the dynamics of the ASEP better than the conventional Burgers equation. }
    \label{fig:CompBurgersASEP2}
  \end{center}
\end{figure}

\begin{figure}[h]
  \begin{center}
    \begin{tabular}{c}
      \begin{minipage}{0.5\hsize}
        \begin{center}
          \includegraphics[height=2.5cm]{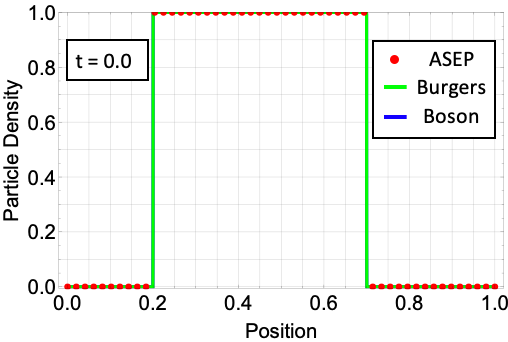}
        \end{center}
      \end{minipage}
      \begin{minipage}{0.5\hsize}
        \begin{center}
          \includegraphics[height=2.5cm]{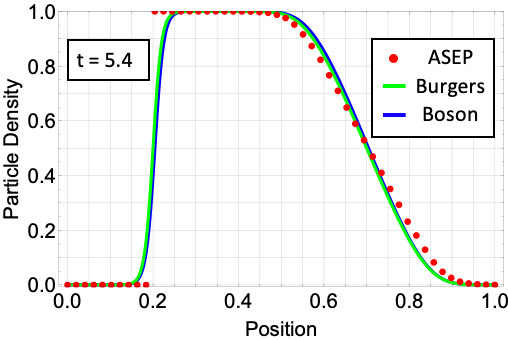}
        \end{center}
      \end{minipage} \\
      
      \begin{minipage}{0.5\hsize}
        \begin{center}
          \includegraphics[height=2.5cm]{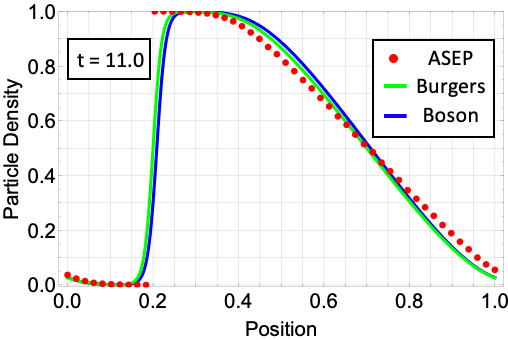}
        \end{center}
      \end{minipage}            
      \begin{minipage}{0.5\hsize}
        \begin{center}
          \includegraphics[height=2.5cm]{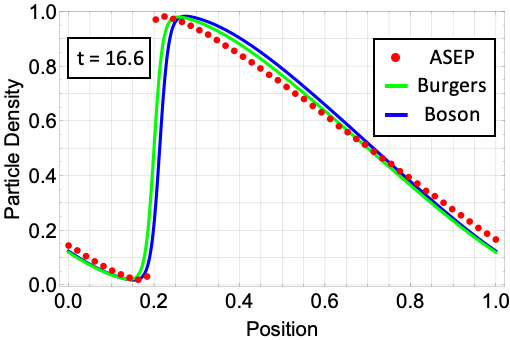}
        \end{center}
      \end{minipage}  \\
      
    \begin{minipage}{0.5\hsize}
        \begin{center}
          \includegraphics[height=2.5cm]{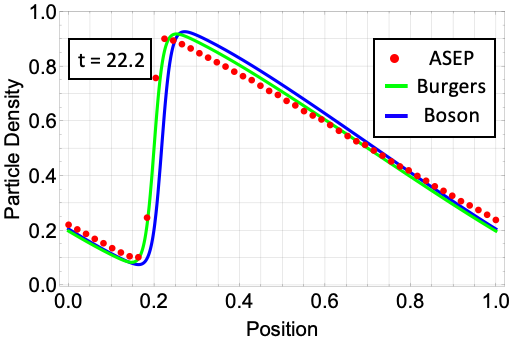}
        \end{center}
      \end{minipage}            
      \begin{minipage}{0.5\hsize}
        \begin{center}
          \includegraphics[height=2.5cm]{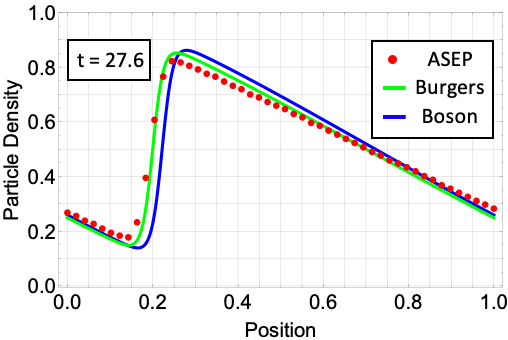}
        \end{center}
      \end{minipage}  \\
      
    \end{tabular}
   \caption{Particle density of the ASEP (red dots), and numerical solutions of the Burgers equation (green lines) and of the Boson-Burgers equation (blue lines) for particle number $N=25$. The receding phenomenon of the solution of the Burgers equation is not observed in this case.}
    \label{fig:CompBurgersASEP3}
  \end{center}
\end{figure}

In order to examine how well this equation describes the dynamics of the ASEP, we compare the numerical solution of the Burgers equation with the particle density profile of the ASEP which is simulated by the Monte Carlo method.
We show the results in Fig. \ref{fig:CompBurgersASEP}, \ref{fig:CompBurgersASEP2}, and \ref{fig:CompBurgersASEP3}.
The particle density of the ASEP is plotted against the position.
Red dots correspond to the ASEP and green lines to the Burgers equation.
Configuration of the simulation of the ASEP is as follows.
The number of lattice $L$ is $50$, and the asymmetric diffusion rate $(p,q)$ is $(1,0)$.
We simulate for various numbers of particles $N$ and show three patterns ($N=1,2,25$) in this paper.
The parameter $a$ of the Burgers equation is determined by the configuration of the ASEP.
We choose the total length of the system as $l=1$ and the lattice distance $a$ as $a=l/L=1/50$.
According to the initial configuration of the ASEP, we set the initial condition of the Burgers equation as the following rectangular function: 
if site $j$ is occupied in the ASEP, then $\rho(x,t)=1$ for $a(j-1) \le x \le aj$, otherwise $0$.
For example, if $N=1$ and a particle is located at site $i$ in the initial state, we denote the initial condition of the Burgers equation as
$\rho_{ini}(x,0)=\theta(x-i)-\theta(i+1-x)$, where $\theta(x)$ is the step function.

As can be seen in Fig. \ref{fig:CompBurgersASEP}, \ref{fig:CompBurgersASEP2} and \ref{fig:CompBurgersASEP3}, the solution of the Burgers equation (\ref{eq:burgers}) roughly describes the dynamics of particle density of the ASEP.
However, Fig. \ref{fig:CompBurgersASEP} and Fig. \ref{fig:CompBurgersASEP2} reveal that the solution of the Burgers equation is receding against the particle density wave of the ASEP. 
This receding phenomenon is observed in various initial conditions, especially with small numbers of particles.
On the other hand, in the case of a large numbers of particles (Fig. \ref{fig:CompBurgersASEP3}), the receding phenomenon is not observed.
Thus, the Burgers equation derived from the conventional way overestimates the exclusive volume effect of the ASEP 
when the number of particles in the ASEP is small. 
The reason for this overestimate is that it considers the arbitrary state at time $t$ in eq. (\ref{eq:state}), which contains states with large numbers of particles in the evaluation of the expectation value.
However, the number of particles of the ASEP is conserved in the periodic boundary condition. 
Therefore, when the number of particles is small in the initial condition, the Burgers equation overestimates the exclusive volume effect.

\textit{The relation between the ASEP and the DNLS.}---
Next we consider the non-exclusive boson model in a one dimensional lattice in which each site can contain more than one particle.
This model is defined by the Hamiltonian
\begin{equation}
\begin{split}
    -\Hboson=&
    \sum_j \hat{a}_j^{\dagger} \left[ p\hat{a}_{j-1}+q\hat{a}_{j+1}-(p+q)\hat{a}_j \right] \\
    &+(p-q) \sum_j \hat{a}_j^{\dagger}(\hat{a}_j^{\dagger}-\hat{a}_{j+1}^{\dagger})\hat{a}_j \hat{a}_{j+1}
\end{split}
\label{eq:HBoson}
\end{equation}
where $\hat{a}_j^{\dagger}$ and $\hat{a}_j$ are the bosonic creation and annihilation operators whose commutation relations are provided by
$[\hat{a}_j,\hat{a}_k]=[\hat{a}_j^{\dagger},\hat{a}_k^{\dagger}]=0,  [\hat{a}_j,\hat{a}_k^{\dagger}]=\delta_{j,k}$.
The time evolution equation of particle density is the same as that of the ASEP \cite{BosonModel}. 
The first term in eq. (\ref{eq:HBoson}) denotes the asymmetric diffusion of particles and the second term mimics the exclusive volume effect of the ASEP.
As in the case of the ASEP, the projection state $\bra s|$ is given by $\bra s| = \bra 0|\text{exp}(\sum_j \hat{a}_j)$, and the expectation value of the physical quantity $\hat{A}$ in the state $\psi(t)$ is provided by $\bra s|\hat{A}|\psi(t)\ket$ in this case.
When the time evolution obeys the imaginary time Schr\"{o}dinger equation which is the same form as eq. (\ref{eq:MasterASEP}), 
the time evolution equation of particle density of this boson model is identical with that of the ASEP.

We take the continuous limit in this model; namely we take the infinitesimal limit of the lattice distance denoted as $a=\frac{l}{L}$, as we did in the previous section.
We replace the boson operaters $\hat{a},\hat{a}^{\dagger}$ with the field operaters $\ps,\pd$ as $\hat{a}_j \rightarrow \sqrt{a}\ps(x_j),\hat{a}_j^{\dagger}\rightarrow \sqrt{a}\pd(x_j)$ where $x_j:=aj$.
Then the Hamiltonian is expressed as
\begin{equation}
\begin{split}
    -\HDNLS=&a^2D\int dx \pd\partial_x^2\ps -2a\alpha \int dx \pd\partial_x\ps \\
    &-2a^2\alpha\int dx \pd\partial_x\pd\ps\ps,
\label{eq:HDNLS}
\end{split}
\end{equation}
where we introduce two parameters $D=\frac{p+q}{2}$ and $\alpha=\frac{p-q}{2}$.
The commutation relations of the field operators $\ps,\pd$ are given by 
$[\ps(x),\ps(y)]=[\pd(x),\pd(y)]=0, [\ps(x),\pd(y)]=\delta(x-y)$.
This Hamiltonian $\HDNLS$ is known as the derivative nonlinear Schr\"{o}dinger type Hamiltonian. 

\textit{Derivation of the Burgers equation from the DNLS.}---
Hereafter we explain the derivation of the Burgers equation from the DNLS type Hamiltonian $\HDNLS$. 
We introduce the orthogonal basis of the states with $N$ particles as 
$|x_1,\cdots,x_N\ket :=\frac{1}{\sqrt{N!}}\pd(x_1)\cdots\pd(x_N)|0\ket$ and denote a state vector as 
\begin{equation}
    |f\ket=a^{\frac{N}{2}}\int dx_1\cdots dx_N f(x_1,\cdots,x_N) |x_1\cdots x_N\ket
\end{equation} 
where $f(x_1,\cdots,x_N)$ is the distribution function of $N$ particles. Then the state vector at time $t$ is given by $|f,t\ket = e^{-\HDNLS t}|f\ket$.

The projection state in the continuous limit is expressed as $|s\ket = e^{\hat{A}}|0\ket$ where $\hat{A}=\int dx a^{-\frac{1}{2}}\pd(x)$.
The commutation relation between $\ps$ and $e^{\hat{A}}$ is $[\ps(x),e^{\hat{A}}]=a^{-\frac{1}{2}} e^{\hat{A}}.$
Letting both sizes of this equation operate on the vacuum $|0\ket$, we obtain
\begin{equation}
    \ps(x)|s\ket=a^{-\frac{1}{2}}|s\ket.
    \label{eq:sProp}
\end{equation}
This relation means that the projection state $|s\ket$ is the eigenvector of the annihilation operator $\ps$ with the eigenvalue $a^{\frac{1}{2}}$.
In addition to this property, the projection state satisfies the following:
\begin{equation}
    \bra s|\HDNLS =0.
\end{equation}
Thus, the projection state is the stationary state of this model.
The expectation value of the physical quantity $\hat{A}$ in the $N$-particles state $|f,t\ket$ is provided by $\bra s|\hat{A}|f,t\ket$.
The time evolution of the physical quantity $\hat{A}(t)$ in the Heisenberg picture is given by 
\begin{equation}
\partial_t \hat{A}=-[\hat{A},\HDNLS].
\end{equation}
Therefore, the time evolution equation of the annihilation operator $\ps$ is described by the quantum DNLS equation:
\begin{equation}
\begin{split}
    \partial_t \ps(x,t) = &a^2D\partial_x^2 \ps(x,t) - 2a\alpha \partial_x \ps(x,t) \\
    &+ 4a^2\alpha \pd(x,t)\ps(x,t) \partial_x \ps(x,t).
\end{split}
\label{eq:qDNLS}
\end{equation}

We denote the density operator as $\hat{\rho}(x)=\pd(x)\ps(x)$. 
Then the particle distribution function at time $t$ is given by
\begin{equation}
\begin{split}
    \rho(x,t)&=\bra s|\hat{\rho}(x)|f,t\ket \\
    &=a^{-\frac{1}{2}}\bra s|\ps(x)|f,t\ket.
\end{split}
\label{eq:disfunc}
\end{equation} 
Here we use the property of the projection state provided by eq. (\ref{eq:sProp}).

Now we are ready to derive the Burgers equation from the quantum DNLS equation. 
Operating the projection state $\bra s|$ from the left and the state vector $|f\ket$ from the right on the quantum DNLS equation (\ref{eq:qDNLS}) and using eq. (\ref{eq:sProp}) and eq. (\ref{eq:disfunc}), then we obtain
\begin{equation}
\begin{split}
        a^{\frac{1}{2}}\partial_t \rho(x,t) =& a^{\frac{5}{2}}D\partial_x^2 \rho(x,t) - 2a^{\frac{3}{2}}\alpha \partial_x \rho(x,t) \\
        &+ 4a^{\frac{3}{2}}\alpha \bra s|\hat{\psi}(x) \partial_x \hat{\psi}(x)|f,t\ket.
\end{split}
\label{eq:beforeApprox}
\end{equation}
In order to calculate the third term on the right hand side, we adopt the mean field approximation, where the distribution function of the $N$-particles state is given by the product of distribution function of the one-particle state:
\begin{equation}
    f(x_1,\cdots,x_N,t)\sim\prod_{j=1}^N f(x_j,t)
\end{equation}
where $f(x,t)$ is the one particle distribution function.
Under this approximation, the particle distribution (\ref{eq:disfunc}) is expressed as 
\begin{equation}
    \rho(x,t)=Nf(x,t).
\end{equation}
Then, the third term on the right hand side of eq. (\ref{eq:beforeApprox}) is calculated as
\begin{equation}
\begin{split}
        \bra s|\hat{\psi}(x) \partial_x \hat{\psi}(x)|f,t\ket  
        &\sim \partial_x\left[\binom{N}{2} f(x,t)^2\right] \\
        &= \frac{N-1}{N}\rho(x,t)\partial_x\rho(x,t).
\end{split}
\label{eq:interactTerm}
\end{equation}
Substituting eq. (\ref{eq:interactTerm}) into eq. (\ref{eq:beforeApprox}), dividing both sides of this equation by $a^{\frac{3}{2}}$ and rescaling time $at \rightarrow t$, one obtains the Burgers equation:
\begin{equation}
    \partial_t \rho = aD\partial_x^2 \rho - 2\alpha \partial_x \rho + 4\alpha \frac{N-1}{N} \rho \partial_x \rho.
\label{eq:bosonBurgers}
\end{equation}
In order to distinguish between the Burgers equation derived from the quantum DNLS equation and the conventional Burgers equation, we call the former equation the  ``Boson-Burgers equation".
The difference between the Boson-Burgers equation and the conventional one is the coefficient of the third term on the right hand side.
The coefficient $\frac{N-1}{N}$ in the Boson-Burgers equation describes the finite particle effect. 
In the limit of $N \to \infty$, the Boson-Burgers equation corresponds to the conventional equation since $\frac{N-1}{N} \to 1$.
As we mentioned above, the conventional Burgers equation overestimates the exclusive volume effect of the ASEP when the particle-number $N$ is small. 
The reason for this deviation is that conventional derivation considers the expectation value among all possible states.
However, in the case of the Boson-Burgers equation, we are able to calculate the expectation value of $N$-particle states in the frame work of the field theory. Therefore we obtain the finite particle correction.

In order to investigate the validity of this finite particle correction, we compare the numerical solutions of the Boson-Burgers equation with the Monte Carlo simulation of the ASEP.
The results are shown in Fig. \ref{fig:CompBurgersASEP}, \ref{fig:CompBurgersASEP2}, and \ref{fig:CompBurgersASEP3}.
The particle density of the ASEP is plotted against position.
The red dots represent the Monte Carlo simulation of the ASEP, while the blue lines represent the numerical solution of the Boson-Burgers equation.
The settings and the parameters of the simulations are the same as in the previous section.

These figures show that the solutions of the Boson-Burgers equation are in good agreement with the particle density of the ASEP. 
For small values of particle number $N$, as in Fig. \ref{fig:CompBurgersASEP} and Fig. \ref{fig:CompBurgersASEP2}, it is observed that the receding phenomena  observed in the solution of the conventional Burgers equation does not appear in the Boson-Burgers equation.

\textit{Conclusion.}---
In this letter, we investigated the relationship between the integrable quantum field theory and the classical integrable systems related to the ASEP.
First, we found that the numerical solutions of the conventional Burgers equation shift backwards against the particle density of the ASEP.
Next, we introduced the non-exclusive boson model, in which the time evolution of the one point correlation function is equivalent to that of the ASEP. The DNLS type Hamiltonian, which is a quantum integrable system, emerged in the continuous limit of this model.
Then we obtained the time evolution equation of the annihilation operator which is the quantum DNLS equation.
Adopting the mean field approximation in the $N$-particle state, we obtained the Boson-Burgers equation, which contains finite particle correlation. We found that this equation describes the dynamics of the ASEP at small numbers of particles better than the conventional Burgers equation.

This work was supported by JST-Mirai Program Grant Number JPMJMI20D1, Japan and JSPS KAKENHI Grant Number JP18K03448.



\end{document}